\documentclass[12pt,preprint]{aastex}

\begin{document}

\slugcomment{accepted to: {\it Icarus}}

\title{Polarization of asteroid (387) Aquitania: the newest member of a class of large inversion angle asteroids}
\author{Joseph Masiero\altaffilmark{1}, Alberto Cellino\altaffilmark{2}}

\altaffiltext{1}{Institute for Astronomy, University of Hawaii, 2680 Woodlawn Dr, Honolulu, HI 96822, {\it masiero@ifa.hawaii.edu}}
\altaffiltext{2}{INAF - Osservatorio Astronomico di Torino, strada Osservatorio 20, I-10025 Pino Torinese (TO), Italy, {\it cellino@oato.inaf.it}}

\begin{abstract}

We present new imaging polarimetric observations of two Main Belt asteroids,
(234) Barbara and (387) Aquitania, taken in the first half of 2008 using the
Dual-Beam Imaging Polarimeter on the University of Hawaii 2.2 meter
telescope, located on Mauna Kea, Hawaii.  Barbara had been previously shown
to exhibit a very unusual polarization-phase curve by \citet{cellino06}.  Our
observations confirm this result and add Aquitania to the growing class of
large inversion angle objects.  Interestingly, these asteroids show spinel
features in their IR spectra suggesting a mineralogical origin to the phase
angle-dependent polarimetric features.  As spinel is associated with
calcium-aluminum-rich inclusions and carbonaceous chondrites, these large
inversion angle asteroids may represent some of the oldest surfaces in the
solar system.  Circular as well as linear polarization measurements were
obtained but circular polarization was not detected.

\end{abstract}


\section{Introduction}
The radiation we receive from asteroids at visible wavelengths is sunlight
scattered by the solid surfaces of the objects. The scattering process
polarizes the emerging photon flux, with the most general state of
polarization being partial elliptical polarization.  In the case of asteroid
scattering, linear polarization dominates over circular and is modulated by
the properties of the surface (e.g. albedo, texture, composition, regolith
size) and the illumination conditions.  Measuring the degree of linear
polarization can diagnose physical conditions of the scattering surface and
is complementary to photometry and spectroscopy for the remote analysis of
small solar system objects.

A common way to quantify the polarization of light is by using the Stokes
vectors $I$, $Q$, $U$, \& $V$, where $I$ is the total intensity, $Q$ the
amount of light polarized in the $0^\circ$ or $90^\circ$ planes, $U$ the
light in the $45^\circ$ or $-45^\circ$ planes, and $V$ the light polarized
circularly with left- or right-handedness.  In particular, Stokes $Q$ and
$U$ fully describe the state of linear polarization.  For cases of linear
polarization, $Q$ and $U$ can be converted into a measure of the total degree
of polarization ($P$) and the position angle of polarization
($\theta$) using: $P = \frac{\sqrt{Q^2+U^2}}{I}$ and $\theta =
\frac{1}{2} \arctan{U/Q}$.

The reference for the position angle of polarization is typically celestial
north, though definitions in the instrument frame or the galactic frame can
also be used if convenient.  In asteroid polarimetry it is found that with
few exceptions the orientation of the plane of linear polarization is
perpendicular or parallel to the scattering plane: the plane encompassing the
Sun, the target, and the observer.  Since most scattering cases result in
light polarized perpendicular to the scattering plane we define that
direction as the reference for polarization position angle.  The new degree
of polarization is then defined as \begin{displaymath} P_r =
\frac{(I_{\perp}-I_{\parallel})}{(I_{\perp}+I_{\parallel})}
\end{displaymath}
where $I_{\perp}$ and $I_{\parallel}$ are the intensities of light
perpendicular and parallel to the scattering plane, respectively.  For this
definition, $P_r$ will be positive for the ``normal'' scattering case, or
negative if the scattered light is polarized in the plane of scattering.
This allows us to refer to polarization of an asteroid as either ``positive''
or ``negative'' and still be completely descriptive.

The primary observable in asteroid polarimetric studies is the change in
percent polarization $P_r$ as a function of phase angle (the angle between
the Sun and the Earth as seen from the asteroid).  Close to zero phase angle
$P_r$ tends to zero, but as the phase angle increases $P_r$ becomes
increasingly negative. This is the so-called branch of negative
polarization. After reaching a maximum negative value usually between
$8^\circ$ and $10^\circ$ phase angle, $P_r$ decreases in absolute value (that
is, becomes depolarized) and reaches zero at the so-called inversion angle
($\alpha_0$) usually between $15^\circ$ and $20^\circ$ phase. The
maximum value of negative polarization is an important parameter, usually
indicated as $P_{min}$, and varies mostly between $-0.5\%$ and $-2.0\%$.
Measurements of $P_{min}$ can be used to calculate albedo \citep{zg76}, and
can be used along with inversion angle as an indication of surface texture
\citep{dz79}.

Beyond the inversion angle $P_r$ becomes re-polarized in the positive sense,
increasing nearly linearly with phase. The slope ($h$) of the branch of
positive polarization for phase angles greater than the inversion angle is
another important parameter, since it is known to be diagnostic of the albedo
of the surface \citep[see, e.g.][and references therein]{cellino99}. The
maximum value of positive polarization is reached at values of phase which
are well beyond the maximum values attainable by main belt asteroids observed
from Earth. Only near-Earth objects are occasionally visible over very large
intervals of phase angle, and the maximum value of positive polarization is
observed in some cases to occur at phase angles of the order of $80$ - $100$
degrees.

\citet{shkuratov94} review the many possible physical models that have been
used to try to explain the negative polarization seen in asteroid phase
curves.  The most promising are the coherent backscattering models, described
in e.g. \citet{muinonen89}, which can explain both the photometric opposition
surge observed for atmosphereless bodies as well as the negative polarization
seen for asteroids at small phase angles.  The polarimetric effect in this
theory has a strong dependence on the size and spacing of the scattering
particles meaning that surface chemistry and mineralogy should play a large
role in determining the polarization as a function of phase.  According to
\citet{muinonen02}, coherent backscattering is the key physical process
generating both these effects.

As part of an extensive campaign to characterize the polarization-phase
curves of over $100$ asteroids \citet{cellino05,cellino06} found an
interesting case in the asteroid (234) Barbara, which was found to exhibit an
unusual polarimetric behavior.  Whereas most asteroids have an inversion
angle of around $20^\circ$ Barbara shows strong negative polarization
($\sim-1\%$) for phase angles larger than $25^\circ$.  The authors proposed
that this strange behavior was due to unusual surface properties, likely the
surface mineralogy driving the rare Ld-type classification.
\citet{gilhutton08} performed a follow-up investigation of other L/Ld-type
asteroids, finding four that show these unusual polarization properties.
While not all L-type asteroids show this effect (e.g. (12) Victoria, see
\citet{cellino05}), both Barbara and (980) Anacostia from
\citet{gilhutton08}'s survey show strong spinel features in their IR spectra
\citep{sunshine07}.  \citet{burbine92} identified strong spinel features in
the IR spectra of both (387) Aquitania and Anacostia, so we investigated
Aquitania to look for the same strange polarization signatures seen in other
spinel-rich L-types.

\section{Observations}
\label{obs}

\begin{figure}
\centering
\includegraphics[angle=-90,scale=0.6]{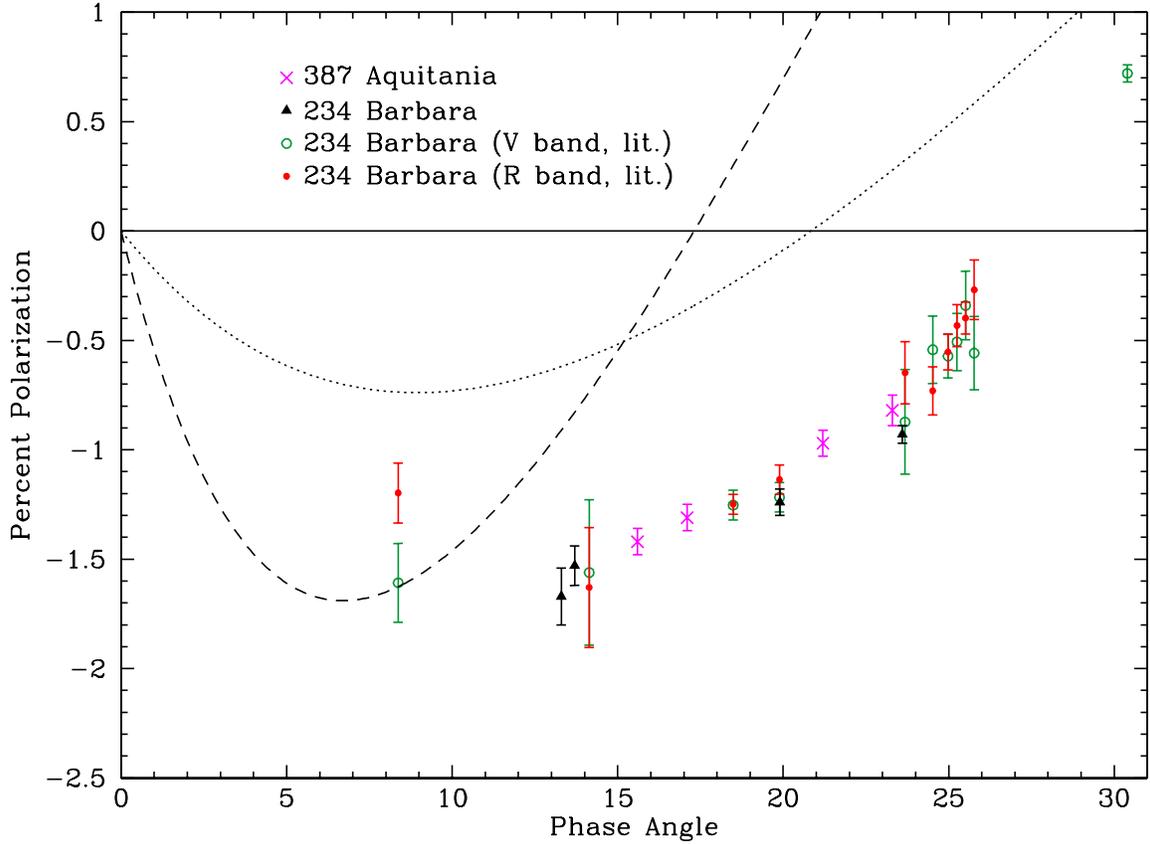}
\protect\caption{
Polarization measurements of (387) Aquitania and (234) Barbara from this work (crosses and filled triangles, respectively) compared to the V and R band Barbara measurements (open and filled circles) from \citet{cellino06}, \citet{cellino07} and \citet{gilhutton08}.  The dashed and dotted lines show typical polarization-phase curves for B-type (e.g. (24) Themis) and L-type (e.g. (12) Victoria) asteroids, respectively \citep{muinonen02, gilhutton08}.}
\label{fig.obs}
\end{figure}

Observations of the asteroids (387) Aquitania and (234) Barbara were
conducted on four nights spread from January to June 2008 on the University
of Hawaii $2.2~$m telescope located on Mauna Kea, Hawaii.  Polarizations were
measured with the Dual-Beam Imaging Polarimeter (DBIP, \citet{dbip}), a
broad-band CCD imaging system sensitive to both linear and circular
polarizations simultaneously.  DBIP has very low instrumental systematics and
can reach polarization precisions of better than $0.1\%$.  DBIP uses an
IR-blocked clear filter which transmits from $400-700~$nm, a close
approximation to a Sloan $g'+r'$ filter.  Although there is some evidence for
color dependence of polarization \citep{cellino05}, the difference in
measured polarization between P$_V$ and P$_R$ is usually within error.  V and
R band polarizations of Barbara showing this consistency can be seen in
Fig~\ref{fig.obs}.  Using a broader filter does not impair our measurements
of the bulk polarization properties.

Stokes parameters were measured for our targets using a beam-swapping pattern
of waveplate alignments.  Each image measured complimentary values of a
single Stokes vector (i.e. $I+Q$ in the north beam and $I-Q$ in the south
beam; $I+U$ north and $I-U$ south; or $I+V$ north and $I-V$ south) which were
then swapped in a second image.  By combining both images we were able to
completely remove time-dependent and flat-field effects on the measured
polarization values.  Individual exposure times were adjusted so that each
image of our target had a peak count value on the CCD between $20,000$ and
$40,000$ counts per pixel to provide enough photons for good statistics while
avoiding non-linear issues near saturation.  Image reduction was completed as
described in \citet{dbip}.

DBIP's native polarization measurement is of the fractional values $Q/I=q$ and
$U/I=u$, which can be converted to $P$ and $\theta$, as well as $V/I=v$.
Table~\ref{tab.obs} presents the asteroid name, UT date of observation,
apparent V magnitude, total exposure time, phase angle ($\alpha$), measured
linear polarization ($P_r$), angle of polarization ($\theta$) and measured
circular polarization.  Total listed exposure time includes all six waveplate
positions needed to develop a full measurement of $q$, $u$, and $v$.  The
linear polarization measurements and angles of polarization have been rotated
into the frame of the asteroid scattering plane.  The circular polarization
values in all cases are consistent with a zero signal to within $2.5\sigma$,
and thus we do not detect any circular polarization of the light scattered
from these asteroids.  Circular polarization is predicted to appear on
asteroids that have surfaces containing powdered metals, especially those with
highly irregular shapes \citep{degtjarev}, however circular polarization as
of yet has not been detected from an asteroid \citep[e.g.,][]{muinonen02}.

Polarized and unpolarized standard star measurements were taken each night in
addition to the asteroid observations.  Standards were drawn from
\citet{fossati07} as well as the standard list published for
Keck/LRISp\footnote{http://www2.keck.hawaii.edu/inst/lris/polarimeter/polarimeter.html}
which includes the {\it Hubble} standards \citep{hubbleSTD2}.  Standard
measurements showed that induced instrumental polarization, systematic
depolarization and instrumental crosstalk were all below the $0.1\%$ level,
and so are not included in the data tables.  See \S~\ref{calib} for further
discussion on systematic error determination.

Our linear polarization data are shown in Figure~\ref{fig.obs} along with the
Barbara data presented in \citet{cellino06}, \citet{cellino07} (these data
also published in \citet{gilhutton08} as the Torino observations), and the
CASPROF data from \citet{gilhutton08}.  In addition, two comparison phase
curves are also plotted.  The first is the fit by \citet{muinonen02} to
polarization data from (24) Themis, a member of the B taxonomic class.
B-type and C-type asteroids typically show the deepest negative polarization
branches.  The other curve is a fit by \citet{gilhutton08} of data for (12)
Victoria, a typical L-class object.  Although different fitting models were
used in these two cases, both models are qualitatively the same at the phase
angles of interest.  Note that Aquitania was classified as an L-class and
Barbara as an Ld-class by \citet{busclass} whereas both are now L-class under
the classification system of \citet{demeo} but show very different
polarization features from other L-class objects.

\section{Instrumental Polarization Calibration}
\label{calib}

Following the calibrations reported in \citet{dbip} a quarterwave retarder
was added in series with the halfwave retarder affecting the instrumental
systematics, especially crosstalk between linear and circular polarization.
To quantify these instrumental errors, extensive lab bench crosstalk
calibrations were performed.  Details of the calibration procedure can be
found in \citet{dbip2}, and are briefly summarized here.

\begin{figure}
\centering
\includegraphics[scale=0.6]{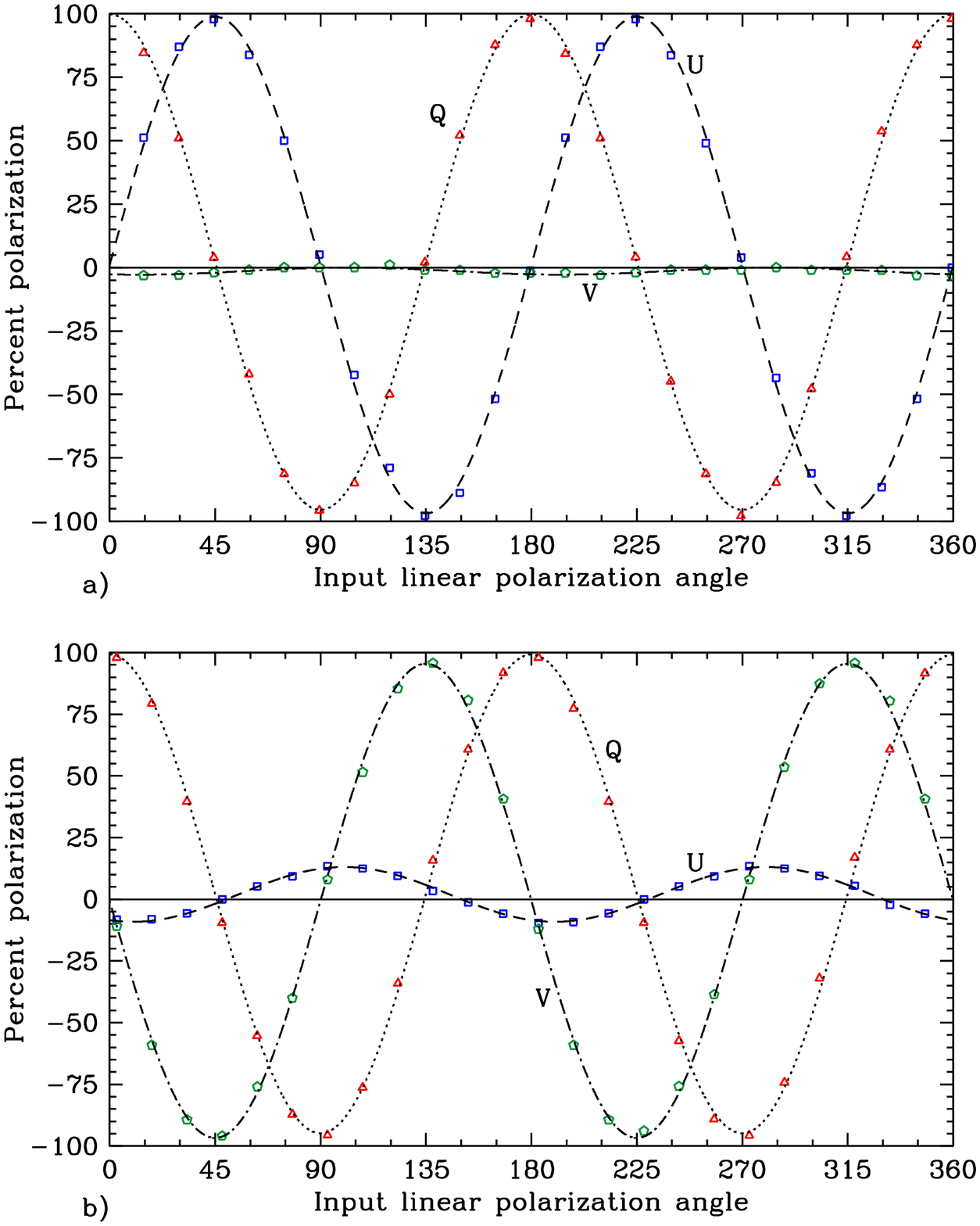}
\protect\caption{
Polarization crosstalk measurements for DBIP for two cases: (a) for
input of pure linear polarization rotated through 360 degrees, and (b) for
input of rotated linear polarization passed through a fixed quarterwave
plate.  Sinusoids for both cases were fitted using a chi-squared minimizer,
and are labeled with the Stokes vectors they represent.  From \citet{dbip2}}
\label{fig.crosstalk}
\end{figure}

Figure~\ref{fig.crosstalk} shows the measured polarization state for an input
of pure linear polarization stepped through $360~$degrees both without and
with a fixed quarterwave plate in the light path (Figs~\ref{fig.crosstalk}a
and \ref{fig.crosstalk}b, respectively).  Ideally the former test should
show pure $Q$ and $U$ polarization as offset sinusoids peaking at $100\%$
with zero $V$ polarization, while the latter test should show offset $Q$ and
$V$ (in this case, the $U$ that is in the same phase as $Q$ indicates a slight
misalignment of the quarterwave plate at the input, while the out of phase
$U$ indicates crosstalk).  A chi-squared minimizer was used to fit the
measured variations in the Stokes vectors to determine crosstalk and
depolarization.  In all cases, the crosstalk and depolarization were found to
be a few percent of the input polarization, so that for sources with
``typical'' polarizations (i.e. $5-10\%$) the errors are comparable to the
desired statistical errors of $\sim0.1\%$ polarization, as has been measured
for polarized and unpolarized standards.

\section{Discussion}
\label{disc}

The new measurements of (234) Barbara presented here agree with those from
\citet{cellino06} and \citet{cellino07}.  Despite very different optical
designs, data acquisition methods, observing circumstances, and physical
locations, the consistency between our results and those from previous work
(e.g. \citet{cellino06}) indicate that comparisons and combinations of
results from these instruments are legitimate.

The similarity between the Barbara and (387) Aquitania polarization values
clearly point toward Aquitania being a member of the new class of large
inversion angle (LIA, or ``Barbarian'') asteroids that includes Barbara,
(172) Baucis, (236) Honoria, (679) Pax and (980) Anacostia
\citep{gilhutton08}.  Barbara, Aquitania, and Anacostia all show strong
$2\mu$ spinel features in their IR spectra \citep{sunshine07, burbine92}
implying a mineralogical origin to the unusual polarization properties of
these asteroids.  The other three LIA objects do not currently have published
IR spectral coverage at $2\mu$, but we predict that they too will show the
same spinel feature.  \citet{sunshine07} explain that a spinel-rich spectrum
can be used as a tracer of Calcium-Aluminum-Rich Inclusions (CAIs), one of
the oldest known materials in the Solar System, as determined from meteorite
chemical analysis (e.g. \citet{sunshine08}).

\citet{gilhutton08} propose that the surfaces of these bodies may be coarse
regoliths of a dark matrix mixed with smaller white inclusions.  This mixing
of two components with different albedos can alter the behavior of bulk
polarization as a function of phase and explain the large inversion angle
seen in these objects.  The authors point out that their sample of LIA
asteroids covers a wide range of semimajor axis-space ($2.38-2.80~$AU) and
thus cannot be fragments from a single parent body.

Table~\ref{tab.ast} shows that Aquitania has orbital and physical properties
fairly similar to those of Anacostia \citep{gilhutton08}.  This similarity
was also pointed out by \citet{burbine92} when they were both identified as
spinel-bearing asteroids.  However, it is very unlikely that these two
objects originate from the same parent body, based on the velocity spread
between their orbits compared to those seen for other disrupted bodies and
dynamical families \citep{willmanPC,zappala95}.  In other words, if both
Aquitania and Anacostia were fragments from the collisional disruption of a
common parent body, the difference in orbital elements would imply
unrealistic ejection velocities of the order of many km/sec. Moreover, given
the non-negligible sizes of Aquitania and Anacostia, such a collisional event
should also be expected to have generated an important dynamical family that
should be possible to identify even after a very long time. Such a family,
however, is simply not found.

It is worth noting that of the six asteroids now identified as LIA objects
four have periods longer than $20$ hours when the average for these sizes is
$\sim10-15~$hours.  Though still a victim of small number statistics, if this
trend of long rotation periods holds for other LIA asteroids it may indicate
that they are characterized by rotation periods significantly longer than the
typical values found for most large Main Belt asteroids \citep{pravecAIII}.
This would imply a highly porous, ``fluffy'' internal structure that is very
efficient at absorbing energy from impacts without transferring it into an
increase in rotation rate.  This kind of structure is similar to what was
observed for the Tagish Lake meteorite \citep{zolensky} and what is expected
for carbonaceous chondrites in which all CAIs found thus far have been
observed \citep{burbine02}.  This would support the theory proposed by
\citet{sunshine08} that spinel-bearing asteroids, and thus probably all LIAs,
are composed of pristine material from the beginning of the solar system.  We
note also that long spin periods might also be diagnostic of binarity,
although any physical reason for a correlation between binarity and anomalous
polarimetric properties remains unknown.

\section{Acknowledgments}

The authors wish to recognize and acknowledge the very significant cultural
role and reverence that the summit on Mauna Kea has always had within the
indigenous Hawaiian community.  We are most fortunate to have the opportunity
to conduct observations from this sacred mountain.  We appreciate the helpful
comments from reviewers R. Gil-Hutton and I. Belskaya which greatly improved
this paper.  We also would like to thank R. Jedicke for editing and
comments on the manuscript.  J.M. was partially supported under NASA PAST
grant NNG06GI46G.  A.C. was supported under ASI contract *I/015/07/0 *.

\begin{deluxetable}{cccccccc}
\tablenum{1}
\tabletypesize{\footnotesize}
\tablecaption{Asteroid Polarization Measurements}
\tablewidth{0pt}
\tablehead{
\colhead{Target}   &
\colhead{UT Obs Date}   &
\colhead{V mag}   &
\colhead{T$_{exp}$ (sec)}   &
\colhead{$\alpha$}  &
\colhead{Linear $\%~$Pol$^a$}  &
\colhead{$\theta_p$}  &
\colhead{Circ $\%~$Pol$^a$}  
}
\startdata
387 Aquitania & 2008-01-17 & 12.7 & 360 & $21.2^\circ$ & $-0.97 \pm 0.06$ & $91.36 \pm 2.57$ & $0.03 \pm 0.02$ \\
 & 2008-03-12 & 11.7 & 72 & $15.6^\circ$ & $-1.42 \pm 0.06$ & $88.95 \pm 1.84$ & $-0.07 \pm 0.04$ \\
 & 2008-05-14 & 11.3 & 72 & $17.1^\circ$ & $-1.31 \pm 0.06$ & $89.58 \pm 1.35$ & $0.00 \pm 0.03 $ \\
 & 2008-06-11 & 11.7 & 60 & $23.3^\circ$ & $-0.82 \pm 0.07$ & $90.16 \pm 2.54$ & $-0.03 \pm 0.04$ \\
234 Barbara & 2008-01-17 & 14.3 & 1620 & $19.9^\circ$ & $-1.24 \pm 0.06$ & $92.11 \pm 2.80$ & $0.06 \pm 0.03$ \\
 & 2008-03-12 & 13.6 & 720 & $23.6^\circ$ & $-0.93 \pm 0.04$ & $89.33 \pm 1.61$ & $0.08 \pm 0.03$ \\
 & 2008-05-14 & 12.2 & 270 & $13.7^\circ$ & $-1.53 \pm 0.09$ & $87.98 \pm 2.70$ & $-0.01 \pm 0.03$ \\
 & 2008-06-11 & 12.0 & 108 & $13.3^\circ$ & $-1.67 \pm 0.13$ & $89.07 \pm 2.45$ & $0.06 \pm 0.03$ \\
\hline
\enddata
\vskip 0.05in
\scriptsize{$^a~$quoted errors are $1\sigma$ statistical errors; systematic errors are $\approx0.05\%$}.
\label{tab.obs}
\end{deluxetable}

\begin{deluxetable}{cccccccc}
\tablenum{2}
\tabletypesize{\footnotesize}
\tablecaption{Orbital and Physical Parameters}
\tablewidth{0pt}
\tablehead{
\colhead{Asteroid}   &
\colhead{a (AU)$^a$}   &
\colhead{e$^a$}   &
\colhead{i (deg)$^a$}  &
\colhead{H (mag)$^b$}  &
\colhead{Diameter (km)$^b$}  &
\colhead{Rot. Period (h)$^c$}  
}
\startdata
387 Aquitania & 2.739 & 0.237 & 18.14 & 7.41 & 100.51 & 24.144 \\
980 Anacostia & 2.743 & 0.200 & 15.90 & 7.85 & 86.19 & 20.117 \\
\hline
\enddata
\vskip 0.05in
\scriptsize{$^a~$from MPCORB: {\it http://www.cfa.harvard.edu/iau/MPCORB.html}.  $^b~$from \citet{tedesco02}.  $^c~$from {\it http://cfa-www.harvard.edu/iau/lists/LightcurveDat.html}}.
\label{tab.ast}
\end{deluxetable}


\begin{thebibliography}{XXX}

\bibitem[Burbine et~al.(2002)]{burbine02}
Burbine, T.H., McCoy, T.J., Meibom, A., Gladman, B. \& Keil, K., 2002, ``Meteoritic Parent Bodies: Their Number and Identification'',  Asteroids III, ed. Bottke, Cellino, Paolicchi \& Binzel (Univ of Arizona Press), 653.

\bibitem[Burbine et~al.(1992)]{burbine92}
Burbine, T.H., Gaffey, M.J. \& Bell, J.F., 1992, ``S-asteroids 387 Aquitania and 980 Anacostia - Possible fragments of the breakup of a spinel-bearing parent body with CO3/CV3 affinities'',  Meteoritics, 27, 424.

\bibitem[Bus \& Binzel(2002)]{busclass}
Bus, S. \& Binzel, R., 2002, ``Phase II of the Small Main-Belt Asteroid Spectroscopic Survey:  A Feature-Based Taxonomy'', Icarus, 158, 146.

\bibitem[Cellino et~al.(2007)]{cellino07}
Cellino, A., di Martino, M., Levasseur-Regourd, A.-C., Belskaya, I.N., Bendjoya, Ph. \& Gil Hutton, R., 2007, ``Asteroid compositions: Some evidence from Polarimetry'', Advances in Geosciences, 7, 21.

\bibitem[Cellino et~al.(2006)]{cellino06}
Cellino, A., Belskaya, I.N., Bendjoya, Ph., di Martino, M., Gil Hutton, R., Muinonen, K. \& Tedesco, E.F., 2006, ``The strange polarimetric behavior of Asteroid (234) Barbara'', Icarus, 180, 565.

\bibitem[Cellino et~al.(2005)]{cellino05}
Cellino, A., Gil Hutton, R., di Martino, M., Bendjoya, Ph., Belskaya, I.N. \& Tedesco, E.F., 2005, ``Asteroid polarimetric observations using the Torino UBVRI photopolarimeter'', Icarus, 179, 304.
	
\bibitem[Cellino et~al.(1999)]{cellino99}
Cellino, A., Gil Hutton, R., Tedesco, E.F., di Martino, M. \& Brunini, A., 1999, ``Polarimetric Observations of Small Asteroids: Preliminary Results'', Icarus, 138, 129.

\bibitem[Degtjarev \& Kolokolova(1992)]{degtjarev}
Degtjarev, V.S. \& Kolokolova, L.O., 1992, ``Possible application of circular polarization for remote sensing of cosmic bodies'', EM\&P, 57, 213.

\bibitem[Demeo(2007)]{demeo}
DeMeo, F.E., 2007, ``DeMeo Taxonomy: Categorization of Asteroids in the Near-Infrared'', S.M. thesis, Massachusetts Institute of Technology.

\bibitem[Dollfus \& Zellner(1979)]{dz79}
Dollfus, A. \& Zellner, B., 1979, ``Optical polarimetry of asteroids and laboratory samples'', Asteroids (Univ of Arizona Press), 170.

\bibitem[Fossati et~al.(2007)]{fossati07}
Fossati, L., Bagnulo, S., Mason, E., Landi Del'Innocenti, E., 2007, ``Standard Stars for Linear Polarization Observed with FORS1'', ASP Conf., 364, 503.

\bibitem[Gil-Hutton et~al.(2008)]{gilhutton08}
Gil-Hutton, R., Mesa, V., Cellino, A., Bendjoya, P., Penaloza, L. \& Lovos, F., 2008, ``New cases of unusual polarimetric behavior in asteroids'', A\&A, 482, 309.

\bibitem[Masiero et~al.(2007)]{dbip}
Masiero, J., Hodapp, K.-W., Harrington, D. \& Lin, H., 2007, ``Commissioning of the Dual-Beam Imaging Polarimeter for the UH 88-inch telescope'', PASP, 119, 1126.

\bibitem[Masiero et~al.(2008)]{dbip2}
Masiero, J., Hodapp, K.-W., Harrington, D. \& Lin, H., 2008, ``Extended Commissioning and Calibration of the Dual-Beam Imaging Polarimeter'', to appear in ASP Conf. Series for Astronomical Polarimetry 2008; arXiv:0809.4313.

\bibitem[Muinonen(1989)]{muinonen89}
Muinonen, K., 1989, ``Electromagnetic Scattering by Two Interacting Dipoles'', Proc. URSI International Symp. on Electromagnetic Theory, 428.

\bibitem[Muinonen et~al.(2002)]{muinonen02}
Muinonen, K., Piironen, J., Shkuratov, Y., Ovcharenko, A. \& Clark, B., 2002, ``Asteroid Photometric and Polarimetric Phase Effects'', Asteroids III, ed. Bottke, Cellino, Paolicchi \& Binzel (Univ of Arizona Press), 123.

\bibitem[Pravec et~al.(2002)]{pravecAIII}
Pravec, P., Harris, A.W. \& Michalowski, T., 2002, ``Asteroid Rotations'', Asteroids III, ed. Bottke, Cellino, Paolicchi \& Binzel (Univ of Arizona Press), 113.

\bibitem[Schmidt et~al.(1992)]{hubbleSTD2}
Schmidt, G.D., Elston, R., \&  Lupie, O.L., 1992, ``The Hubble Space Telescope Northern-Hemisphere Grid of Stellar Polarimetric Standards'', AJ, 104, 1563.

\bibitem[Shkuratov et~al.(1994)]{shkuratov94}
Shkuratov, Yu.G., Muinonen, K., Bowell, E., Lumme, K., Peltoniemi, J.I., Kreslavsky, M.A., Stankevich, D.G., Tishkovetz, V.P., Opanasenko, N.V., \& Melkumova, L.Y., 1994, ``A Critical Review of Theoretical Models of Negatively Polarized Light Scattered by Atmosphereless Solar System Bodies'', EM\&P, 65, 201.

\bibitem[Sunshine et~al.(2007)]{sunshine07}
Sunshine, J.M., Connolly, H.C., McCoy, T.J., Bus, S.J., La Croix, L., 2007, ``Identification of Refractory-rich Asteroids: Evidence for the Earliest Accreted Bodies in the Solar System'', LPI, 38, 1613.

\bibitem[Sunshine et~al.(2008)]{sunshine08}
Sunshine, J.M., Connolly, H.C., McCoy, T.J., Bus, S.J., La Croix, L.M., 2008, ``Ancient Asteroids Enriched in Refractory Inclusions'', Science, 320, 514.

\bibitem[Tedesco et~al.(2002)]{tedesco02}
Tedesco, E.F., Noah, P.V., Noah, M. \& Price, S.D., 2002, ``The Supplemental IRAS Minor Planet Survey'', AJ, 123, 1056.

\bibitem[Willman(2008)]{willmanPC}
Willman, M., 2008, Private Communication.

\bibitem[Zappala et~al.(1995)]{zappala95}
Zappala, V., Bendjoya, Ph., Cellino, A., Farinella, P. \& Froeschle, C., 1995, ``Asteroid families: Search of a 12,487-asteroid sample using two different clustering techniques'', Icarus, 116, 291.

\bibitem[Zellner \& Gradie(1976)]{zg76}
Zellner, B. \& Gradie, J., 1976 ``Minor planets and related objects. XX - Polarimetric evidence for the albedos and compositions of 94 asteroids'', AJ, 81, 262.

\bibitem[Zolensky et~al.(2002)]{zolensky}
Zolensky, M.E., Nakamura, K., Gounelle, M., Mikouchi, T., Kasama, T., Tachikawa, O. \& Tonui, E., 2002, ``Mineralogy of Tagish Lake: An ungrouped type 2 carbonaceous chondrite'', M\&PS, 37, 737.

\end{thebibliography}
\end{document}